\documentclass[11pt]{article}
\usepackage[utf8]{inputenc}
\usepackage{enumerate}
\usepackage{xcolor}
\usepackage{amsmath}
\usepackage{amsfonts}
\usepackage{amssymb}
\usepackage{capt-of}
\usepackage[a4paper]{geometry}
\usepackage{float}
\usepackage{graphicx}
\usepackage{authblk}
\usepackage{graphics, setspace}
\usepackage[title]{appendix}
\usepackage{hyperref}
\usepackage{enumitem}
\newcommand{\mathsym}[1]{{}}
\newcommand{\unicode}[1]{{}}
\usepackage{cite}

\begin{document}

\providecommand{\abs}[1]{\lvert#1\rvert}
\newcommand\ii{\'{\i}}
\newcommand\oo{\'{\o}}
\title{New vacuum boundary effects of massive field theories\\}
\author[1]{Manuel Asorey}
\author[1]{Fernando Ezquerro}
\author[1]{Miguel Pardina}

\affil[1]{\ Departamento de F\'isica Te\'orica, Centro de Astropart\'{\i}culas y F\'{\i}sica de Altas Energ\'{\i}as, Universidad de Zaragoza, 50009 Zaragoza, Spain}

\date{}
\maketitle

\begin{abstract}
	\normalsize
	
	Analytical arguments suggest that the Casimir energy in 2+1 dimensions for gauge theories exponentially decays with the distance between the boundaries. The phenomenon  has also been observed by non-perturbative numerical simulations.  The dependence of this exponential decay on the different boundary conditions could help to better understand  the infrared behavior of these theories and in particular their mass spectrum. A similar behavior is expected  in 3+1 dimensions. 
Motivated by this feature we analyze the dependence of the exponential decay of  Casimir energy for different  boundary conditions of massive scalar fields in 3+1 dimensional spacetimes. 
We show that the boundary conditions classify in two different families according on the rate of this exponential decay of the Casimir energy. If the boundary conditions  on each boundary are independent (e.g. both boundaries satisfy Dirichlet boundary conditions), the Casimir energy has a exponential decay that is two times faster than when the boundary conditions interconnect  the two boundary plates (e.g. for periodic or antiperiodic boundary conditions).  
These results will be useful for a comparison with the Casimir energy in the non-perturbative regime of non-Abelian gauge theories. 
\end{abstract}
\providecommand{\keywords}[1]
{
	\small	
	\textbf{Keywords:} #1
}

\keywords{Casimir energy; non-abelian gauge theories, boundary conditions in field theory}
\newpage
\section{Introduction}

Boundary effects of quantum fields play a crucial role in different quantum phenomena. The Casimir effect is one of the first and clearest examples  of these boundary effects \cite{Casimir:1948dh}. In this case, the renormalized energy of the vacuum becomes dependent of the boundary conditions when the quantum field is confined between two parallel plates. The variation in vacuum energy with the distance between the two plates gives rise to a force between them, which is determined by the specific conditions satisfied by the quantum fields at the boundaries. Despite of its very tiny magnitude the effect has been observed for massless fields in several different experimental configurations \cite{sparnaay1957attractive,sparnaay1958measurements,lamoreaux1997demonstration,PhysRevLett.81.4549,PhysRevLett.87.211801, Dalvit,PhysRevLett.88.041804}.

Significant progress regarding the computation and understanding of the Casimir effect in different models and configurations has been made in recent years. Notable findings were presented in Ref. \cite{Boundary_general_2013}, in which the vacuum energy for massless scalar fields under very general boundary conditions was obtained in arbitrary dimensions using analytic spectral representation techniques. Later on, the  temperature dependence was computed for 
3+1 dimensional massless field theories in Ref. \cite{munoz2020thermal}. More recently, the massive case was studied in the low temperature limit  for general boundary conditions in 2+1 dimensional space-times \cite{Casimir2+1}.

However, the Casimir effect in interacting theories is much less known \cite{SYMANZIK19811}. Recently, there has been some progress in the analytic approach to a non-perturbative calculation of  the Casimir energy for Yang-Mills theories in 2+1 dimensions \cite{karabali1996origin,KARABALI1998661,PhysRevD.98.105009}. This computation has been later corroborated by some numerical simulations using Dirichlet boundary conditions for $SU(2)$ gauge fields \cite{PhysRevLett.121.191601}.

This analytic approach is based on the parametrization of  $SU(2)$ gauge field by means of a massive scalar field, whose (magnetic) mass  $m={g^2}/{\pi}$ is given in terms of the gauge coupling $g$. For Dirichlet boundary conditions the
non-perturbative Casimir energy derived from numerical simulations does agree with that of a massive scalar field with 
such a magnetic mass.

To clarify whether or not there exists a similar relation between the Casimir energies of 3+1 dimensional  Yang-Mills theories and  3+1 dimensional  massive scalar fields \cite{asorey2023energy}, some numerical simulations are in progress. Some preliminary results using Dirichlet boundary conditions can been found in Ref. \cite{Sim3+1}.  In order to make a more comprehensive  comparison with the results of a massive scalar field it is necessary first to understand how the Casimir energy behaves under  different boundary conditions.

 Since comparisons with lattice gauge theory results require working at finite temperature, it is crucial to study the effects of thermal fluctuations on Casimir energy at low temperatures in massive scalar field theories, to provide a more accurate analytical reference for the results. 

The paper is organized as follows. In section \ref{sectRenorm} we introduce the set up and the renormalization scheme to be used throughout the paper. In section \ref{sectionLowT} we develop  the formalism needed to obtain the  free energy in the low temperature regime. We make use of these results in section \ref{sectionCasimir}  to compute the Casimir energy, and also, to explore its decay in the large distance limit. In section \ref{sectionparticular} we analyze  some particular boundary conditions and compare their behavior. The conclusions of the paper are summarized in section \ref{Conclusions}, and  finally, we include in Appendix \ref{app1} the computation of the Casimir energy using the explicit form of the spacial eigenvalues for those particular boundary conditions.

\section{Renormalized effective action}\label{sectRenorm}

Let us consider a free massive scalar field $\phi$ confined between two homogenous plates separated  a distance $L$.
In the Euclidean formalism  the effects of finite temperature can be described by a compactification of Euclidean time into a circle of radius $\frac{\beta}{2\pi}=\frac1{2\pi T}$ with periodic boundary conditions  $\phi(t+\beta,\mathbf{x})=\phi(t,\mathbf{x})$. The corresponding partition function of the quantum field  is
\begin{equation}\label{det}
	Z(\beta)=\text{det}\left(-\partial_0^2-\boldsymbol{\nabla}^2+m^2\right)^{-1/2},
\end{equation}
where $\partial_0$ denotes the Euclidean time derivative, $\boldsymbol{\nabla}^2$ the spacial Laplacian and $m$ the mass of the fields. Since the infinite boundary plates are homogeneous we can describe the boundary conditions by means 
of $2\times 2$ unitary matrices $U$ \cite{asorey2005global} 
\begin{equation}
	\psi - i\delta\dot\psi = U(\psi +i\delta\dot\psi );\hspace{1cm}U\in \text{U(2)},
	\label{bccc}
\end{equation}
where 
\begin{equation}
	\psi=\begin{pmatrix}
		\psi(L/2)  \\ \psi(-L/2)  \\ 
	\end{pmatrix},
	\quad
	\dot\psi=\begin{pmatrix}
		\dot\psi(L/2)  \\ \dot\psi(-L/2)  \\ 
	\end{pmatrix},
\end{equation}
 $\psi(\pm L/2)= \phi(t,x_1,x_2, \pm  L/2)$ being the values of  
the fields  $\phi$ at the boundary plates,  $\dot\psi(\pm L/2)=\pm\partial_3(\phi(t,x_1,x_2, \pm L/2))$  the outward normal  derivatives on the plates, and $\ell$  is an arbitrary scale parameter that we shall set  $\ell=1$ for simplicity.

The $U(2)$ matrices can be parametrized in the following way
\begin{eqnarray}\label{U2}
	U(\theta,\eta,{{\bf n}})&=&{\mathrm e}^{i\theta}\left(\mathbb{I }\cos\eta+i{{\bf n}}\cdot\boldsymbol{\sigma}\,\sin\eta \right);
	\label{parametrization}\quad  {\theta\in[0,2\pi),\,\, \eta\in[-\pi/2,\pi/2)},
\end{eqnarray}
 where  $\boldsymbol{\sigma}$  are the Pauli matrices and $\bf n$ is a three dimensional unit vector ${{\bf n}}\in S^2$. These parameters are restricted to the domain \textcolor{black}{$0\leq\theta\pm\eta\leq\pi$} in order to keep the operator  $-\boldsymbol{\nabla}^2$ selfadjoint and non-negative \cite{Boundary_general_2013}. Also, we have to impose that $n_2=0$ because the scalar field we are working with is real.

The determinant in equation (\ref{det}) 
is ultraviolet divergent. We will regularize that determinant by using the zeta function regularization method \cite{PhysRevD.13.3224,Blau:1988kv}. The regularized effective action  is 
\begin{equation}
	S_{\text{eff}}=-\log Z=-\frac{1}{2}\frac{d}{ds}\zeta \left(s\right)|_{s=0},
\end{equation}
where
\begin{equation}\label{zeta_first}
	\zeta(\beta,m,L;s)=\text{tr} \left(\mu^{2s}\left(m^2-\boldsymbol{\nabla}^2-\partial_0^2\right)^{-s}\right).
\end{equation}
and
$\mu$ is a scale  parameter that makes the zeta function dimensionless and encodes the standard renormalization group parametrization \cite{PhysRevD.56.7797,10.1063/1.532929}. The renormalization prescription consists in fixing this parameter by physical constrains.

 The eigenvalues of the operator $-\square=-\partial_0^2-\boldsymbol{\nabla}^2$  in the current set up are given by the sum of the two dimensional  longitudinal continuous spacial modes $q^2$ that are parallel to the plates, the discrete spacial modes $q_j$ on the transverse direction to the plates that depend on the boundary conditions, and the 
 square of the temporal modes $(2\pi l/\beta)^2$ which are related to the Matsubara frequencies, i.e.
\begin{equation}
	\lambda= \left(\frac{2\pi l}{\beta}\right)^2+q^2+q_j^2+m^2 \hspace{2cm}j\in {\mathbb N}, l \in \mathbb Z.
\end{equation} 
Hence, we can express the zeta function \eqref{zeta_first} in the following form
\begin{equation}\label{zeta_3d}
	\zeta(\beta,m,L;s)=\mu^{2s}\frac{S}{4\pi (s-1)}\sum_{l=-\infty}^\infty\sum_{j} \left(\left(\frac{2\pi l}{\beta}\right)^2+q_j^2+m^2\right)^{-s+1} .
\end{equation}
where $S$ denotes the area of the boundary plates and we have integrated out the continuous spacial modes. A trivial analysis shows that the zeros of the following spectral function  \cite{Boundary_general_2013}
\begin{equation}\label{spectral_3d}
	h^L_U(q)=2i\left(\sin(qL)\left((q^2-1)\cos \eta +(q^2+1)\cos \theta \right)-2q\sin \theta \cos (qL)-2qn_1\sin \eta\right).
\end{equation}
give the discrete spacial modes.
Thus, by using Cauchy theory we can write the zeta function as a contour integral 
\begin{equation}\label{zeta_integral}
	\zeta(\beta,m,L;s)=\mu^{2s}\frac{S}{8\pi i(s-1)}\sum_{l=-\infty}^\infty \oint dq \left(\left(\frac{2\pi l}{\beta}\right)^2+q^2+m^2\right)^{-s+1}\frac{d}{dq}\log h^L_U(q)
\end{equation}
along a thin loop enclosing all the zeros of the spectral function $h_U(k)$ that are in the positive real axis.

This expression has UV divergences that come from the zero temperature contributions and have the following asymptotic expansion when $L$ is large \cite{Boundary_general_2013,asorey_temp1}
\begin{equation}
	S_{\text{eff}}^{l=0}= \beta E_0 =
	C_v(m)S\beta L+C_b(m)S\beta+S\beta\ C_c(m,L)+\ldots. 
\end{equation}
 where $C_b(m)$ is the vacuum energy density on the boundary plates which is divergent, $C_v(m)$ the bulk vacuum energy density that is also divergent and $C_c(m,L)$ is  the finite coefficient corresponding to the finite Casimir energy.

For removing these divergences we need a renormalization prescription with a physical meaning. Our choice is to define the renormalized action as  \cite{asorey_temp1, asorey2015topological} 
\begin{equation}\label{zeta_combination}
	S^{\text{ren}}_{\text{eff}}=-
	\frac{1}{2}\frac{d}{ds} \zeta_{\text{ren}}
	(\beta,m,L;s) 
	\Bigr|_{s=0}\ ,
\end{equation}
where
\begin{equation}\label{zeta_re}
	\zeta_{\text{ren}}(\beta,m,L;s)=\lim_{{L_0\rightarrow \infty}}\left(\zeta(\beta,m,L;s)+\zeta(\beta,m,L+2L_0;s)-2\zeta(\beta,m,L+L_0;s)\right),
\end{equation}
and we have used an auxiliary length $L_0$. This renormalization scheme not only removes the divergences, but also the residual terms that are independent of  the distance between the plates or have a linear dependence on it. The remaining finite part consists of the Casimir energy and some extra terms that do not depend linearly in $\beta$ and also vanish as the distance between the plates   goes to infinity. That is precisely the physical condition we use to fix the renormalization scheme prescription.

\section{Zeta function in low temperature regime}\label{sectionLowT}

We can explicitly compute the sum of the Matsubara modes  in the low temperature regime, i.e. $\beta m\gg  1$, postponing  the analysis of contribution of the spacial modes.  The expression \eqref{zeta_3d} can be written as
\begin{equation}
	\zeta(\beta,m,L;s)=\left(\frac{\beta\mu}{2\pi}\right)^{2s}\frac{ \pi S}{\beta^2(s-1)}\sum_{j}\sum_{l=-\infty}^\infty \left( l^2+\left(\frac{q_j \beta}{2\pi}\right)^2+\left(\frac{\beta m}{2\pi}\right)^2\right)^{-s+1}.
\end{equation}
and applying Mellin transform and  Poisson formula for the temperature dependent modes we get
\begin{equation}\label{Poisson}
	\zeta(\beta,m,L;s)=\left(\frac{\beta\mu}{2\pi}\right)^{2s}\frac{ \pi^{3/2}S}{\Gamma(s)\beta^2}\sum_{j}\sum_{l=-\infty}^\infty\int_0^\infty dt\ t^{s-5/2}\	e^{-\left(\left(\frac{q_j \beta}{2\pi}\right)^2+\left(\frac{\beta m}{2\pi}\right)^2\right)t-\frac{(\pi l)^2}{t}}.
\end{equation}
The integral can be easily computed giving as a result
\begin{align}\nonumber
	\zeta(\beta,m,L;s)=&\left(\frac{\beta\mu}{2\pi}\right)^{2s}\frac{\pi^{3/2} S}{\Gamma(s)\beta^2}\left(\Gamma \left(s-\frac{3}{2}\right)\sum_i\left(\left(\frac{q_j \beta}{2\pi}\right)^2+\left(\frac{\beta m}{2\pi}\right)^2\right)^{3/2-s}\right.\\ 
	\nonumber
	&\left. +4\sum_{j}\sum_{l=1}^\infty\left(\pi l\right)^{-3/2+s}\left(\left(\frac{q_j \beta}{2\pi}\right)^2+\left(\frac{\beta m}{2\pi}\right)^2\right)^{3/4-s/2}K_{3/2-s}\left(\beta l\sqrt{q_j^2+m^2}\right)\right),
\end{align}
where the first term is linear on $\beta$ and is the leading one in the zero temperature limit ($l=0$), whereas the rest ($l\not =0$) have a non-linear dependence on $\beta$. 

Let us first  focus on the leading term. The sum of the boundary modes can be replaced by an integral modulated by the spectral function \eqref{spectral_3d}
\begin{equation}
	\zeta^{l=0}(\beta,m,L;s)=\mu^{2s}\frac{S\beta  \Gamma \left(s-3/2\right)}{16\pi^{5/2} i \Gamma(s)}\oint dq\ \left(q^2+m^2\right)^{3/2-s}\frac{d}{dq}\log h^L_U(q).
\end{equation}
The renormalized zeta  function in our renormalization scheme \eqref{zeta_re} is
\begin{equation}\nonumber 
	\zeta_{\text{ren}}^{l=0}(\beta,m,L;s)=\frac{\mu^{2s} S\beta  \Gamma \left(s-3/2\right)}{16\pi^{5/2} i \Gamma(s)}\lim_{L_0\rightarrow \infty}\oint dq \left(q^2+m^2\right)^{3/2-s}\frac{d}{dq}\log \frac{h^L_U(q)h^{2L_0+L}_U(q)}{\left(h^{L_0+L}_U(q)\right)^2}.
\end{equation}
which is  free of UV divergences. Thus, the only divergent contribution left in the zeta function is the $\Gamma(s)$ factor of the denominator, which disappears when 
computing
the derivative of $\zeta_{\text{ren}}^{l=0}(\beta,m,L;s)$ with respect to $s$ and evaluating it at $s=0$, 
\begin{equation}
	\left(\zeta_{\text{ren}}^{l=0}\right)'(\beta,m,L;0)=\frac{S \beta }{12\pi^2 i }\lim_{L_0\rightarrow \infty}\oint dq \left(q^2+m^2\right)^{3/2}\left(\frac{d}{dq}\log \frac{h^L_U(q)h^{2L_0+L}_U(q)}{\left(h^{L_0+L}_U(q)\right)^2}\right).
\end{equation}
Now, the loop integral can be simply reduced to an integral over the imaginary axis by using the fact that the integrand is holomorphic
\begin{equation}
		\left(\zeta_{\text{ren}}^{l=0}\right)'(\beta,m,L;0)=-\frac{S \beta }{12\pi^2 i }\lim_{L_0\rightarrow \infty}\int_{-\infty}^\infty dq \left(m^2-q^2\right)^{3/2}\left(\frac{d}{dq}\log \frac{h^L_U(iq)h^{2L_0+L}_U(iq)}{\left(h^{L_0+L}_U(iq)\right)^2}\right).
\end{equation}

Notice, that since the integrand is parity odd, the integral between $(-m,m)$  vanishes, but the branching point of the square root $\sqrt{m^2-k^2}$ at $-m$ introduces a change of sign and thus the contributions between $(-\infty,-m)$ and $(m,\infty)$ have the same sign. Taking  this into account the integral reduces to
\begin{equation}
	\left(\zeta_{\text{ren}}^{l=0}\right)'(\beta,m,L;0)=\frac{S \beta }{6\pi^2 }\lim_{L_0\rightarrow \infty}\int_{m}^\infty dq \left(q^2-m^2\right)^{3/2}\left(\frac{d}{dq}\log \frac{h^L_U(iq)h^{2L_0+L}_U(iq)}{\left(h^{L_0+L}_U(iq)\right)^2}\right).
\end{equation}
Finally, we can take the limit $L_0\rightarrow\infty$ in the spectral functions by noticing that 
\begin{equation}\label{L0 infinity}
	\lim_{L_\ast\rightarrow \infty} h_U^{L_\ast}(iq)=\lim_{L_\ast\rightarrow \infty}e^{q(L_\ast)}\left((q^2+1)\cos \eta +(q^2-1)\cos \theta +2q\sin\theta \right),
\end{equation}
and the result can be expressed in terms of this limit  
\begin{equation}
	h^\infty_U(iq)\equiv \left((q^2+1)\cos \eta +(q^2-1)\cos \theta +2q\sin\theta \right),
\end{equation} which leads to the simpler expression
\begin{equation}
	\left(\zeta_{\text{ren}}^{l=0}\right)'(\beta,m,L;0)=-\frac{S\beta }{6\pi^2 }\int_{m}^\infty  dq \left(q^2-m^2\right)^{3/2}\left(L-\frac{d}{dq}\log \frac{h^L_U(iq)}{h^{\infty}_U(iq)}\right).
\end{equation}

\subsection{Contribution of the modes with $l\not =0$}
In this case 
\begin{align} \nonumber
	\zeta^{l\not =0}(\beta,m,L;s)=&\left(\frac{\beta\mu}{2\pi}\right)^{2s}\frac{4\pi^{3/2}S}{\Gamma(s)\beta^2}\sum_{j}\sum_{l=1}^\infty\left(\pi l\right)^{-3/2+s}\\
	&\times  \left(\left(\frac{q_j \beta}{2\pi}\right)^2+\left(\frac{\beta m}{2\pi}\right)^2\right)^{3/4-s/2}
	K_{3/2-s}\left(\beta l\sqrt{q_j^2+m^2}\right)
\end{align}
 the sums over the two series are convergent since the Bessel function $K_{3/2}$ exponentially decays
as its  argument grows. Therefore, only the factor $\Gamma(s)$  is divergent which after derivation and evaluation at $s=0$  gives
\begin{equation}
	\left(\zeta^{l\not =0}\right)'(\beta,m,L;0)=\frac{S}{\pi\beta^2 }\sum_j\left(\beta \sqrt{q_j^2+m^2}\ \text{Li}_2\left(e^{-\sqrt{q_j^2+m^2}\beta}\right)+\text{Li}_3\left(e^{-\sqrt{q_j^2+m^2}\beta}\right)\right).
\end{equation}
Once again, we can use the spectral formula \eqref{spectral_3d} for the sum of the discrete modes $q_j$
\begin{align} \nonumber
	\left(\zeta^{l\not =0}\right)'(\beta,m,L;0)
	=&\frac{S}{2\pi^2 i\beta^2}\oint dq\ \left(\beta \sqrt{q^2+m^2}\ \text{Li}_2\left(e^{-\sqrt{q^2+m^2}\beta}\right)\right.\\
	&\left.+\text{Li}_3\left(e^{-\sqrt{q^2+m^2}\beta}\right)\right) \frac{d}{dq}\log\left(h^L_U(q)\right),
\end{align}
which gives rise to the renormalized zeta function \eqref{zeta_re}
\begin{align} \nonumber
	\left(\zeta^{l\not =0}_{\text{ren}}\right)'(\beta,m,L;0)
	=&\lim_{L_0\rightarrow \infty}\frac{S}{2\pi^2 i\beta^2}\oint dq\ \left(\beta \sqrt{q^2+m^2}\ \text{Li}_2\left(e^{-\sqrt{q^2+m^2}\beta}\right)\right.\\
	&\left.+\text{Li}_3\left(e^{-\sqrt{q^2+m^2}\beta}\right)\right) \frac{d}{dq}\log \frac{h^L_U(q)h^{2L_0+L}_U(q)}{\left(h^{L_0+L}_U(q)\right)^2}.
\end{align}
Following similar steps as in the case  $l=0$  we can reduce this integral to the imaginary axis because the integrand is holomorphic
\begin{align} \nonumber
	\left(\zeta^{l\not =0}_{\text{ren}}\right)'(\beta,m,L;0)=&-\lim_{L_0\rightarrow \infty}\frac{S}{2\pi^2i \beta^2}\int_{-\infty}^\infty dq\ \left(\beta \sqrt{m^2-q^2}\ \text{Li}_2\left(e^{-\sqrt{m^2-q^2}\beta}\right)\right.\\
	&\left.+\text{Li}_3\left(e^{-\sqrt{m^2-q^2}\beta}\right)\right) 
	\left(\frac{d}{dq}\log \frac{h^L_U(iq)h^{2L_0+L}_U(iq)}{\left(h^{L_0+L}_U(iq)\right)^2}\right).
\end{align}
Now, since the integrand is odd, the integral cancels between $(-m,m)$. But the branching point of the square root $\sqrt{m^2-q^2}$ adds a minus between the intervals $(-\infty,-m)$ and $(m,\infty)$ on the square root, which, taking into account that Li$_s(z^*)=\text {Li}^*_s(z)$ the imaginary part of the Li$_2$ and  the real part of Li$_3$ vanish. In summary, the integral is reduced to
\begin{align} \nonumber
	\left(\zeta^{l\not =0}_{\text{ren}}\right)'(\beta,m,L;0)=&-\lim_{L_0\rightarrow \infty}\frac{S}{\pi^2 \beta^2}\!\!\int_{m}^\infty\!\!\!dq\! \left(\beta \sqrt{q^2-m^2}\ \Re\left(\text{Li}_2\left(e^{-i\sqrt{q^2-m^2}\beta}\right)\right)\right.\\
	&\left.+\Im \left(\text{Li}_3\left(e^{-i\sqrt{q^2-m^2}\beta}\right)\right)\right) 
	 \left(\frac{d}{dq}\log \frac{h^L_U(iq)h^{2L_0+L}_U(iq)}{\left(h^{L_0+L}_U(iq)\right)^2}\right),
\end{align}
and using equation \eqref{L0 infinity} in the limit $L_0\rightarrow\infty$  we get
\begin{align} \nonumber
	\left(\zeta^{l\not =0}_{\text{ren}}\right)'(\beta,m,L;0)=&\frac{S}{\pi^2 \beta^2}\!\!\int_{m}^\infty\!\!\!dq\!  \left(\beta \sqrt{q^2-m^2}\ \Re\left(\text{Li}_2\left(e^{-i\sqrt{q^2-m^2}\beta}\right)\right)\right.\\
	&\left.+\Im \left(\text{Li}_3\left(e^{-i\sqrt{q^2-m^2}\beta}\right)\right)\right) 
	\left(L-\frac{d}{dq}\log \frac{h^L_U(iq)}{h_U^\infty(iq)}\right).
\end{align}

\section{Free energy}\label{sectionCasimir}

From the renormalized zeta function  we can easily obtain the Casimir energy. The temperature independent part of the free energy $F=S_{\text{eff}}/\beta$ is the Casimir energy \cite{bordag2018free} 
\begin{equation}\label{Cas_spectral}
	F^{l=0}_U(\beta,m,L)=E_U^c(m,L)=\frac{S }{12\pi^2 }\int_{m}^\infty  dq \left(q^2-m^2\right)^{3/2}\left(L-\frac{d}{dq}\log \frac{h^L_U(iq)}{h^{\infty}_U(iq)}\right),
\end{equation}
whereas the $l\not=0$ terms encode the temperature dependence contributions
\begin{align}\nonumber
	F^{l\not =0}_U(\beta,m,L)=&-\frac{S}{2\pi^2 \beta^3}\!\!\int_{m}^\infty\!\!\!dq\! \left(\beta \sqrt{q^2-m^2}\ \Re\left(\text{Li}_2\left(e^{-i\sqrt{q^2-m^2}\beta}\right)\right)\right.\\
	&\left.+\Im \left(\text{Li}_3\left(e^{-i\sqrt{q^2-m^2}\beta}\right)\right)\right) \left(L-\frac{d}{dq}\log \frac{h^L_U(iq)}{h_U^\infty(iq)}\right).\label{Free_spectral}
\end{align}
Both contributions to the free energy vanish when the distance between the plates $L$ tends to infinity, which is the physical requirement of the renormalization scheme prescription chosen. Moreover, in the zero temperature limit  $F^{l\not =0}$ vanishes and we recover the zero temperature energy.

\subsection{Casimir energy in the asymptotic limit }
Let us now explore the behavior of the Casimir energy in the limit when  the effective distance $mL$ tends to infinity $mL\rightarrow \infty$. First, we can rewrite the spectral function as
\begin{equation}\nonumber
	h_U^L(iq)=e^{qL}\left((q^2+1)\cos\eta+(q^2-1)\sin\theta+2q\sin\theta\right)\left(1+n_1\sin(\eta) {\mathcal {X}}\ e^{-qL}+{\mathcal {Y}}\ e^{-2qL}\right),
\end{equation}
where ${\mathcal {X}}$ and ${\mathcal {Y}}$ are
\begin{align}
	&{\mathcal {X}}(q,\theta,\eta)=\frac{4q}{(q^2+1)\cos\eta+(q^2-1)\sin\theta+2q\sin\theta}\\
	&{\mathcal {Y}}(q,\theta,\eta)=\frac{-(q^2+1)\cos\eta-(q^2-1)\sin\theta+2q\sin\theta}{(q^2+1)\cos\eta+(q^2-1)\sin\theta+2q\sin\theta}.
\end{align}
Using  the asymptotic expansion in powers of $e^{-qL}$ of the quotient of spectral functions 
\begin{equation} 
	\log \frac{h_U^L(iq)}{h_U^\infty(iq)}=qL+n_1\sin \eta\,
	 {\mathcal {X}}\ e^{-qL}+({\mathcal {Y}}-{{\mathcal {X}}’}/{2})e^{-2qL}+O(e^{-3qL}),
\end{equation}
an introducing this expansion in the integral of the Casimir energy we get
\begin{align}\nonumber
	E_U^c&=-\frac{S}{12\pi^2}\int_m^\infty dq (q^2-m^2)^{3/2}\frac{d}{dq}\left(n_1\sin\eta\ {\mathcal {X}}\ e^{-qL}+({\mathcal {Y}}-{{\mathcal {X}}’}/{2})e^{-2qL}+O(e^{-3qL})\right)\\
	&=\frac{S}{4\pi}\int_m^\infty dq\ q(q^2-m^2)^{1/2}\left(n_1\sin\eta {\mathcal {X}}\ e^{-qL}+({\mathcal {Y}}-{{\mathcal {X}}’}/{2})e^{-2qL}+O(e^{-3qL})\right),
\end{align}
where ${\mathcal {X}}'=(n_1\sin(\eta){\mathcal {X}})^2$.
This integral can be estimated by using the saddle point approximation for each exponential order
\begin{equation}
	\int_{m}^{\infty}e^{F(x)}dx\simeq e^{F(x_0)}\int_{m}^{\infty}e^{\frac{1}{2}F''(x_0)(x-x_0)^2}\ dx
\end{equation}
where we consider the quadratic approximation of  $F(x)$ around it maximum at $x_0$. In the limit  $mL\rightarrow\infty$, the maximum is attained in the integration domain at the point $x_0=m(1+a/(mL))$, $a$ being a positive numerical parameter. This means we can express the integral of each exponential order as
\begin{equation}\nonumber
	\int_m^\infty dq\ g(\theta,\eta,n_1,q)e^{-jqL}=\frac{e^{-jmL}}{(mL)^{3/2}}\left(c_{j,1}(\theta,\eta,n_1,m)+\frac{c_{j,2}(\theta,\eta,n_1,m)}{mL}+O\left(\frac{1}{(mL)^2}\right)\right).
\end{equation}
Then the Casimir energy formula can be written as
\begin{align}
	E_U^c=&\frac{Sm^{3}}{(mL)^{3/2}}\left(n_1\sin\eta\  e^{-mL}\left(c_{1,1}(\theta,\eta,n_1,m)+\frac{c_{1,2}(\theta,\eta,n_1,m)}{mL}+O\left(\frac{1}{(mL)^2}\right)\right)\right.\\
	&\left.+e^{-2mL}\left(c_{2,1}(\theta,\eta,n_1,m)+\frac{c_{2,2}(\theta,\eta,n_1,m)}{mL}+O\left(\frac{1}{(mL)^2}\right)\right)+O(e^{-3mL})\right).
\end{align}

The coefficient of the leading exponential term $e^{-mL}$ vanishes when $n_1\sin\eta=0$,  which increases the Casimir energy  exponential rate decay by a factor 2, $e^{-2mL}$. Thus, we have two distinct families of boundary conditions according to the asymptotic behavior of the Casimir energy
\begin{equation}\label{rate}
	(mL)^{3/2} E_U^c\sim \left\{ \begin{matrix}e^{-mL} &\text{if  }\text{ tr}(U\sigma_1)\not =0\\
		e^{-2mL} &	\text{if  }\text{ tr}(U\sigma_1)=0,  \end{matrix} \right.
\end{equation}
which is determined by the dependence or not on $\sigma_1$ of the the unitary matrix $U$ that parametrizes the boundary conditions. 

The classification of the boundary conditions into two different families according to the speed of the exponential decay of Casimir energy for a massive scalar field in 3+1 dimensions, is the main result of this paper. A similar classification also appears in 2+1 dimensions \cite{Casimir2+1}.

These two families are differentiated in whether $\text{tr}(U\sigma_1)$ is zero or not. When $\text{tr}(U\sigma_1)$ does not vanish, the boundary conditions involve a relation between the boundary values or the normal derivatives at the plates. In contrast, when $\text{tr}(U\sigma_1)$ is zero, the constraints  imposed by the boundary conditions at  the normal derivatives and the values at the boundary plates are independent one from each other.

This result was found previously for Dirichlet  and periodic boundary conditions  \cite{cougo1994schwinger}--
\cite{Fulling2005MassDO}, but we have
found that it is a more general property and has a physical meaning. In fact, the same result can be proven for any space dimension $D$.

$\,$
\bigskip

\section{ Particular cases of boundary conditions}\label{sectionparticular}

We can compute the  Casimir energy for some special boundary conditions that are of interest because they can also be implemented for gauge fields. An alternative calculation for some of these boundary conditions 
is carried out in Appendix \ref{app1} by using the exact spectrum of the spacial Laplacian instead of the spectral function.

\begin{enumerate}[leftmargin=*,label={},leftmargin=0pt]
    \item i) \textit{Periodic boundary conditions}:  $\psi(L/2)=\psi(-L/2)$ and $\dot\psi(L/2)=-\dot\psi(-L/2)$. The associated unitary operator is $U_P=\sigma_1$. The logarithm of spectral functions has the form
    \begin{equation}
    	\frac{d}{dq}\log\left(h^L_{U_P}(iq)/h^\infty_{U_P}(iq)\right)=L\coth (qL/2).
    \end{equation}
    The integral of the Casimir energy is
    \begin{align}\label{Cas_Per}
    	E^c_{P}(L,m)=-\frac{Sm^2}{2\pi^2L}\sum_{j=1}^\infty \frac{K_2(jmL)}{j^2},
    \end{align}
    when $m\not = 0$, and
    \begin{equation}
    	E^c_P(L,0)=-\frac{\pi^2S}{90 L^3}
    \end{equation}
in the massless case. In the large $mL$ limit this expression decays exponentially as $e^{-mL}$ which is the result expected since the unitary matrix $U_P$ depends on $\sigma_1$, i.e. $\mathrm{Tr}\, \sigma_1 \,\sigma_1= 2\neq 0$. The temperature dependent terms of the free energy are
    \begin{align}\nonumber
    	F^{l\not =0}_{P}(\beta,m,L)=&-\frac{SL}{2\pi^2 \beta^3}\int_{m}^\infty dq\ \left(\beta \sqrt{q^2-m^2}\ \Re\left(\text{Li}_2\left(e^{-i\sqrt{q^2-m^2}\beta}\right)\right)\right.\\
    	&\left.+\Im \left(\text{Li}_3\left(e^{-i\sqrt{q^2-m^2}\beta}\right)\right)\right) \left(1-\coth (qL/2)\right). \label{Free_Per}
    \end{align}
    Although this integral cannot be analytically computed, the asymptotic expansion of $1-\coth (qL/2)$ shows that it has the same asymptotic behavior as the Casimir energy
    
	\item ii) \textit{Dirichlet boundary conditions}: $\psi(L/2)=\psi(-L/2)=0$. The corresponding  unitary matrix is $U_D=-\mathbb{I}$ and the spectral function  is 
	\begin{equation}\label{spectra_dir}
		\frac{d}{dq}\log\left(h^L_{U_D}(iq)/h^\infty_{U_D}(iq)\right)=L\coth (qL),
	\end{equation}
and the Casimir energy is given by 
\begin{align}\label{Cas_Dir}
	E^c_{D}(m,L)=-\frac{Sm^2}{8\pi^2L}\sum_{j=1}^\infty \frac{K_2(2jmL)}{j^2},
\end{align}
when $m\not =0$, whereas for the massless case we have
\begin{equation}
	E^c_D(0,L)=-\frac{\pi^2S}{1440 L^3}.
\end{equation}
Since $-\mathrm{Tr}\, \mathbb{I}\, \sigma_1= 0$, we have the expected asymptotic behaviour where the Casimir energy decays as $e^{-2 mL}$.
For $l\not =0$ terms of the free energy we have 
\begin{align}\nonumber
	F^{l\not =0}_{D}(\beta,m,L)=&-\frac{SL}{2\pi^2 \beta^3}\int_{m}^\infty dq\ \left(\beta \sqrt{q^2-m^2}\ \Re\left(\text{Li}_2\left(e^{-i\sqrt{q^2-m^2}\beta}\right)\right)\right.\\
	&\left.+\Im \left(\text{Li}_3\left(e^{-i\sqrt{q^2-m^2}\beta}\right)\right)\right) \left(1-\coth (qL)\right)\label{Free_Dir}.
\end{align}
This integral cannot be explicitly analytically computed but it can be shown that decays exponentially in the same way as the Casimir energy.

	\item iii)\textit{ Neumann boundary conditions}: $\dot\psi(L/2)=\dot\psi(-L/2)=0$. Although the corresponding unitary matrix is $U_N=\mathbb I$ and the spectral function is different from the one with Dirichlet boundary conditions the derivative of the logarithm of both spectral functions is the same as \eqref{spectra_dir}. Thus, the free energy is the same as for Dirichlet boundary conditions.
	
	\item iv) \textit{Anti-periodic boundary conditions}:   $\psi(L/2)=-\psi(-L/2)$ and $\dot\psi(L/2)=\dot\psi(-L/2)$. The
	corresponding unitary matrix is $U_A=-\sigma_1$ and in this case the derivative of the logarithm  of the spectral function is
	\begin{equation}
		\frac{d}{dq}\log\left(h^L_{U_A}(iq)/h^\infty_{U_A}(iq)\right)=L\tanh (qL/2).
	\end{equation}
	Thus, the Casimir energy is
	\begin{align}\label{Cas_APer}
		E^c_{U_A}(L,m)=\frac{Sm^2}{2\pi^2L}\sum_{j=1}^\infty \frac{(-1)^{j+1}}{j^2}K_2(jmL)
	\end{align}
	for $m\not = 0$, and  
	\begin{equation}
		E^c_A(L,0)=\frac{7\pi^2S}{720 L^3}
	\end{equation}
	in the massless case. We can see how it has the same asymptotic decay as the periodic boundary conditions, since $U_A$ also depends on $\sigma_1$. The temperature dependent part of the free energy is
	\begin{align} \nonumber
		F^{l\not =0}_{U_A}(\beta,m,L)=&-\frac{SL}{2\pi^2 \beta^3}\int_{m}^\infty dq\ \left(\beta \sqrt{q^2-m^2}\ \Re\left(\text{Li}_2\left(e^{-i\sqrt{q^2-m^2}\beta}\right)\right)\right.\\
		&\left.+\Im \left(\text{Li}_3\left(e^{-i\sqrt{q^2-m^2}\beta}\right)\right)\right) \left(1-\tanh (qL/2)\right).\label{Free_APer}
	\end{align}
	with  the same exponential decay than the Casimir energy.
	\item v) \textit{Zaremba boundary conditions}. This boundary conditions consists of one boundary wall imposing Neumann conditions meanwhile the other imposes Dirichlet boundary conditions. They are described by the operator $U_Z=\pm \sigma_3$. The derivative of the logarithm of spectral functions is given by
	\begin{equation}
		\frac{d}{dq}\log\left(h^L_{U_Z}(iq)/h^\infty_{U_Z}(iq)\right)=L\tanh (qL).
	\end{equation}
	
	The Casimir energy  is
	\begin{align}\label{Cas_Zar}
		E^c_{U_Z}(L,m)=\frac{Sm^2}{8\pi^2L}\sum_{j=1}^\infty \frac{(-1)^{j+1}}{j^2}K_2(2jmL),
	\end{align}
	when the mass is non-null, whereas 
	\begin{equation}
		E^c_Z(L,0)=\frac{7\pi^2S}{11520 L^3}
	\end{equation}
	for the massless case. The exponential decay is the same as for Dirichlet or Neumann boundary 
	conditions $e^{-2mL}$. The rest of the terms of the free energy have the form
	
	\begin{align}\nonumber
		F^{l\not =0}_{Z}(\beta,m,L)=&-\frac{SL}{2\pi^2 \beta^3}\int_{m}^\infty dq\ \left(\beta \sqrt{q^2-m^2}\ \Re\left(\text{Li}_2\left(e^{-i\sqrt{q^2-m^2}\beta}\right)\right)\right.\\
		&\left.+\Im \left(\text{Li}_3\left(e^{-i\sqrt{q^2-m^2}\beta}\right)\right)\right) \left(1-\tanh (qL)\right).\label{Free_Zar}
	\end{align}
\end{enumerate}

\subsection{Asymptotic behavior of Casimir energy}
\begin{figure}[H]
\hspace{1cm}\includegraphics[width=.8\textwidth]{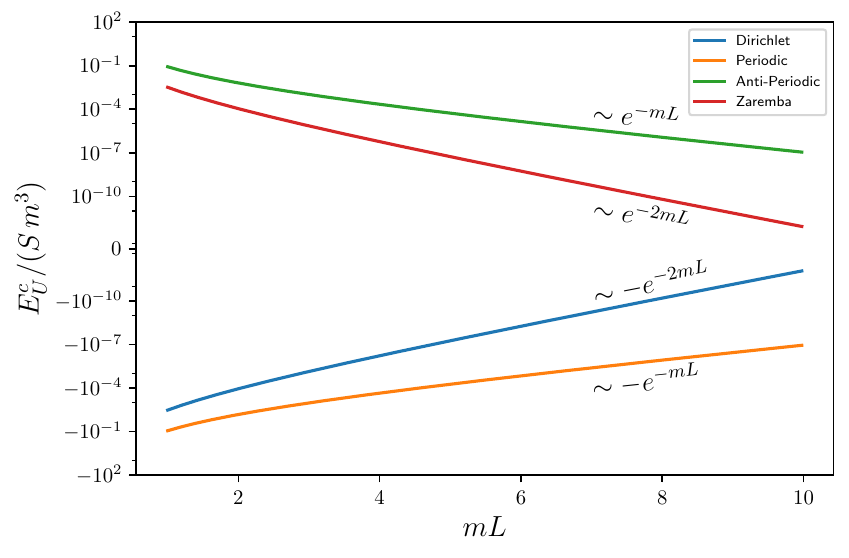}
	\captionof{figure}{Behavior of the adimensional Casimir energy (in logarithmic scale) for  different boundary conditions as a function of the adimensional distance $mL$.}
	\label{plot_Cas}
\end{figure}

From the previous results we can see how the asymptotic behaviour with $mL$ of the Casimir energy  follows a common rule \eqref{rate}; for Neumann, Dirichlet \eqref{Cas_Dir} and Zaremba \eqref{Cas_Zar} boundary conditions it is  exponentially decaying 
\begin{equation}
	(mL)^{3/2} E^c_U\sim e^{-2mL}
\end{equation} 
at the same rate,
which agrees with the prescription given by the general rule because in all these cases
$\text{tr}\left(U\sigma_1\right)=0$. However, the boundary conditions satisfying that  $\text{tr}\left(U\sigma_1\right)\not=0$ like periodic \eqref{Cas_Per} and anti-periodic \eqref{Cas_APer} the exponential decay is
\begin{equation}
	(mL)^{3/2} E^c_U\sim e^{-mL}.
\end{equation} 
The behavior of the  Casimir energy for these  boundary conditions showing  the different asymptotic behaviors is displayed in Figure~\ref{plot_Cas}.

\begin{figure}[H]
\hspace{1cm}	\includegraphics[width=.8\textwidth]{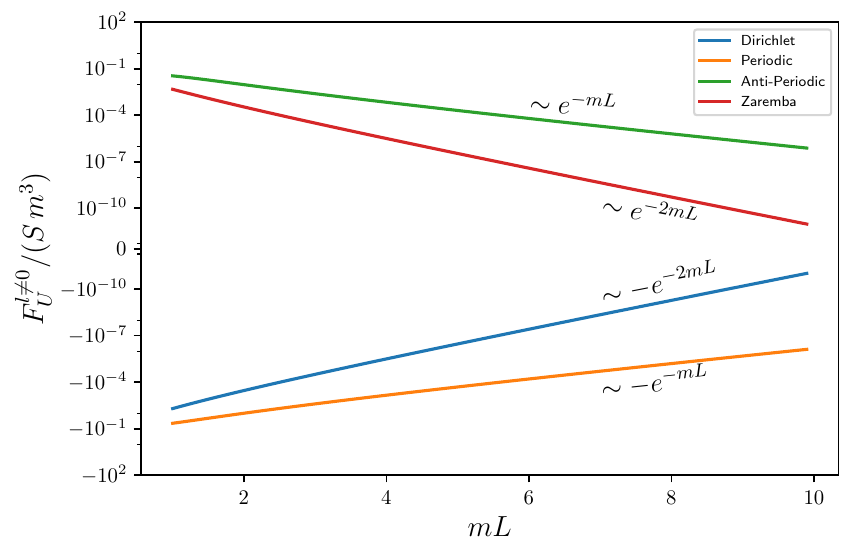}
	\captionof{figure}{Behavior of the adimensional free energy in logarithmic scale for four different boundary conditions
	 as a function of the effective distance $mL$ with a fixed temperature $m\beta=1$. 
	}
	\label{plot_Free}
\end{figure}

On the other hand, we can also plot the rest of the terms of the free energy $F^{l\not=0}_U$, that could not be analytically calculated, to show how they have the same exponential decay with $mL$ as the Casimir energy (see Figure~\ref{plot_Free}).

How the boundary conditions are imposed at each wall is what physically differentiates the two families with different asymptotic behavior. In the family with a faster decay (Dirichlet, Neumann and Zaremba) the boundary conditions on each plate are imposed independently, whereas for the second family (periodic, antiperiodic) there is a connection between the boundary values or the derivatives at the plates, which induces a slower decay.

\section{Conclusions}\label{Conclusions}

In this paper we have shown that the rate of exponential decay at large distances in the Casimir energy of a massive scalar field in 3+1 dimensions, splits the boundary conditions into two different families. These two types differ by the fact that whether or not there is a relationship between the values or the derivatives at the boundary plates.

Although measuring this effect for massive fields seems hopeless because this exponential decay makes it negligeable in comparison with the effect for electrodynamics fields (that experience just a potential decay with the distance), from a theoretical standpoint it can be relevant for studying if the infrared behavior in non-abelian gauge theories can be described by massive scalar fields.

In fact, in 2+1 dimensions numerical simulations \cite{PhysRevLett.121.191601} and some analytic arguments \cite{KARABALI1998661,PhysRevD.98.105009} point out that with Dirichlet boundary conditions the gauge theories have an exponential decay with a mass lower than the lightest glueball. Finding some similar behavior for 3+1 dimensions would provides new hints for a better understanding the infrared regime of non-abelian gauge theories, and in particular the mass-gap problem. In particular, the identification of these two different regimes of exponential decays for the Casimir energy  would provide a very strong support of this conjecture and its implications. Some accurate numerical simulations to test the conjecture  are in progress.

\renewcommand{\theequation}{A\arabic{equation}}
\setcounter{equation}{0}
\begin{appendices}

\section{Explicit calculations of the free energy}\label{app1}
In this section we calculate the free energies of the theory for different boundary conditions (Dirichlet and periodic)   directly by using the known spectrum of the spacial Laplacian   and compare with the results we obtained by using the general method based on the spectral function and the use of Cauchy theorems.

\subsection{Dirichlet boundary conditions}\label{Dir_modes}

When imposing Dirichlet boundary conditions the eigenvalues of the discrete modes are of the form $k_n=\pi n/L$ where $n=1,\ldots,\infty$.

With this eigenvalues  the zeta function \eqref{zeta_3d} is given by
\begin{equation}
	\zeta(\beta,m,L;s)=\left(\frac{\beta\mu}{2\pi}\right)^{2s}\frac{ \pi S }{ \beta^2 (s-1)}\sum_{j=1}^\infty \sum_{l=-\infty}^\infty \left(l^2+\left(\frac{j\beta}{2L}\right)^2+\left(\frac{m\beta}{2\pi}\right)^2\right)^{-s+1}, 
\end{equation}
and operating in the same way as in the general case we can arrive at
\begin{align}  \nonumber
	\zeta(\beta,m,L;s)&=\left(\frac{\beta\mu}{2\pi}\right)^{2s}\frac{\pi^{3/2}S}{\Gamma(s)\beta^2}\left(\Gamma \left(s-\frac{3}{2}\right)\sum_{j=1}^\infty\left(\left(\frac{j \beta}{2L}\right)^2+\left(\frac{\beta m}{2\pi}\right)^2\right)^{3/2-s}\right.\\ \nonumber
	&\left. +4\sum_{j,l=1}^\infty\left(\pi l\right)^{s-3/2}\left(\left(\frac{j \beta}{2L}\right)^2+\left(\frac{\beta m}{2\pi}\right)^2\right)^{3/4-s/2}K_{3/2-s}\left(\beta l\sqrt{\left(\frac{\pi j}{ L}\right)^2+m^2}\right)\right).
\end{align}

Although this expression diverges in the limit $s\to 0$ its first derivative does not.
This can be achieved by first deriving the $\Gamma(s)$ on the $l\not=0$ terms
\begin{align}\nonumber
	\left(\zeta^{l\not=0}\right)'(\beta,m,L;0)=&\frac{S}{\pi\beta^2}\sum_{j=1}^\infty\left(\frac{\beta}{L}\sqrt{(j\pi)^2+(mL)^2}\ \text{Li}_2\left(e^{-\frac{\beta}{L}\sqrt{(j\pi)^2+(mL)^2}}\right)\right.\\
	&+\left.\text{Li}_3\left(e^{-\frac{\beta}{L}\sqrt{(j\pi)^2+(mL)^2}}\right)\right), \label{Temp_Dir}
\end{align}
whereas on the $l=0$ terms we get
\begin{align} \nonumber
	\zeta^{l=0}(\beta,m,L;s)=&\left(\frac{\mu L}{\pi}\right)^{2s}\frac{\pi^2 S\beta}{16L^3 \Gamma(s)}\left(\Gamma \left(s-2\right)\left(\frac{mL}{\pi}\right)^{4-2s}-\frac{\Gamma(s-\frac{3}{2})}{\sqrt \pi }\left(\frac{mL}{\pi}\right)^{3-2s}\right.\\ 
	 & \left.+4\sum_{j=1}^\infty \left(\frac{mL}{j\pi^2}\right)^{2-s} K_{2-s}\left(2jmL\right)  \right),
\end{align}

and by using the Gamma function $\Gamma(s)$ properties we get 
\begin{equation}\label{Dir_f}
	\left(\zeta^{l=0}\right)'(\beta,m,L;0)=\frac{SL\beta m^4}{32\pi^2}\left(2\log\frac{\mu}{m}+\frac{3}{2}\right)-\frac{Sm^3\beta}{12\pi}+\frac{S\beta m^2}{4\pi^2L}\sum_{j=1}^\infty\frac{K_{2}\left(2jmL\right)}{j^2}.
\end{equation}

The Casimir energy can be computed from this expression  by
 using the renormalization prescription we described in \eqref{zeta_combination} 
\begin{equation}
	E^c_D(L,m)=-\frac{Sm^2}{8\pi^2L}\sum_{j=1}^\infty\frac{K_{2}\left(2jmL\right)}{j^2},
\end{equation}
which is the same that we obtained in \eqref{Cas_Dir} by the   spectral function method. In
 the massless case it reduces to
\begin{equation}
	E^c_D(L,0)=-\frac{\pi^2S}{1440L^3}.
\end{equation}
The temperature dependent component \eqref{Temp_Dir}  
\begin{align}\nonumber
	F^{l\not =0}_D(\beta,m,L)\!=&\frac{-S}{2\pi\beta^3}\sum_{j=1}^\infty\left(\frac{\beta}{L}\sqrt{(j\pi)^2\!+\!(mL)^2}\ \text{Li}_2\left(e^{-\frac{\beta}{L}\sqrt{(j\pi)^2+(mL)^2}}\right)\!+\!\text{Li}_3\left(e^{-\frac{\beta}{L}\sqrt{(j\pi)^2+(mL)^2}}\right)\right)\\\nonumber
	&-\frac{1}{2\beta}\lim_{L_0\rightarrow\infty}\left(\left(\zeta^{l\not =0}\right)'(\beta,m,L+2L_0;0)-2\left(\zeta^{l\not =0}\right)'(\beta,m,L+L_0;0)\right),
\end{align}
gives the same result as the that obtained from the general expression \eqref{Free_Dir}.

\subsection{Periodic boundary conditions}\label{Per_modes}

With  periodic boundary conditions, we have  as discrete eigenvalues $k_n=2\pi n/L$  with $j\in \mathbb{Z}$, and the corresponding zeta function 

\begin{equation}
	\zeta(\beta,m,L;s)=\left(\frac{\beta\mu}{2\pi}\right)^{2s}\frac{ \pi S }{ \beta^2 (s-1)} \sum_{j,l=-\infty}^\infty \left(l^2+\left(\frac{j\beta}{L}\right)^2+\left(\frac{m\beta}{2\pi}\right)^2\right)^{-s+1}
\end{equation}
becomes   
\begin{align}  \nonumber
	\zeta(\beta,m,L;s)=&\left(\frac{\beta\mu}{2\pi}\right)^{2s}\frac{\pi^{3/2}S}{\Gamma(s)\beta^2}\left(\Gamma \left(s-\frac{3}{2}\right)\sum_{j=-\infty}^\infty\left(\left(\frac{j \beta}{L}\right)^2+\left(\frac{\beta m}{2\pi}\right)^2\right)^{3/2-s}\right.\\ \nonumber
	& +4\!\sum_{j=-\infty}^\infty\!\sum_{l=1}^\infty\!\left(\pi l\right)^{s-3/2}\left(\left(\frac{j \beta}{L}\right)^2+\left(\frac{\beta m}{2\pi}\right)^2\right)^{3/4-s/2}\\
	&\left. \times K_{3/2-s}\left(\!\beta l\sqrt{\left(\frac{2\pi j}{ L}\right)^2+m^2}\right)\right),
\end{align}
after some simple operations

The $l\not =0$ terms we can just derivate the $\Gamma(s)$ and evaluate in $s=0$
\begin{align}\nonumber
\left(\zeta^{l\not =0}\right)^\prime(\beta,m,L;0)=&\frac{S}{\pi\beta^2}\!\!\sum_{j=-\infty}^\infty\left(\frac{\beta}{L}\sqrt{(j\pi)^2+(mL)^2}\ \text{Li}_2\left(e^{-\frac{\beta}{L}\sqrt{(j\pi)^2+(mL)^2}}\right)\right.\\
&\left.+\text{Li}_3\left(e^{-\frac{\beta}{L}\sqrt{(j\pi)^2+(mL)^2}}\right)\right).\label{Temp_Per}
\end{align}
The $l=0$ term can be rewritten as
\begin{equation}\label{Per1}
	\zeta^{l=0}(\beta,m,L;s)=\left(\frac{\mu L}{2\pi}\right)^{2s}\frac{\pi^{3/2}S\beta \Gamma \left(s-\frac{3}{2}\right)}{8L^3 \Gamma(s-1)(s-1)}\sum_{j=-\infty}^\infty\left(j^2+\left(\frac{m L}{2\pi}\right)^2\right)^{3/2-s},
\end{equation}
and by using the properties of Gamma function $\Gamma(s)$ we get
\begin{equation}\label{Per_f}
	\left(\zeta^{l=0}\right)'(\beta,m,L;0)=\frac{SL\beta m^4}{32\pi^2}\left(\log\frac{\mu}{m}+\frac{3}{4}\right)+\frac{S\beta m^2}{\pi^2L}\sum_{j=1}^\infty\frac{K_{2}\left(jmL\right)}{j^2}.
\end{equation}
Thus,  under the renormalization prescription \eqref{zeta_combination} the Casimir energy is
\begin{equation}
	E^c_P(L,m)=-\frac{Sm^2}{2\pi^2L}\sum_{j=1}^\infty\frac{K_{2}\left(jmL\right)}{j^2},
\end{equation}
 which does coincide with \eqref{Cas_Per}, and in the massless case becomes
\begin{equation}
	E^c_P(L,0)=-\frac{\pi^2S}{90L^3}.
\end{equation}
The temperature dependent component \eqref{Temp_Per} is
\begin{align}\nonumber
	F^{l\not =0}_P(\beta,m,L)=&-\frac{S}{2\pi\beta^3}	\sum_{j=1}^\infty\left(\frac{\beta}{L}\sqrt{(2j\pi)^2+(mL)^2}\ \text{Li}_2\left(e^{-\frac{\beta}{L}\sqrt{(2j\pi)^2+(mL)^2}}\right)\right.\\
	&\left.+\text{Li}_3\left(e^{-\frac{\beta}{L}\sqrt{(2j\pi)^2+(mL)^2}}\right)\right) \nonumber\\ 
	&-\frac{1}{2\beta}\lim_{L_0\rightarrow\infty}\left(\left(\zeta^{l\not =0}\right)'(\beta,m,L+2L_0;0)-2\left(\zeta^{l\not =0}\right)'(\beta,m,L+L_0;0)\right),
\end{align} 
which also does  agree with expression \eqref{Free_Per}.
\end{appendices}

\newpage
\section*{Acknowledgements}
We are partially supported by Spanish Grants No. PGC2022-126078NB-C21 funded by\\ MCIN/AEI/10.13039/ 501100011033, ERDF A way of making EuropeGrant; the Quantum Spain project of the QUANTUM ENIA of Ministerio de Economía y Transformación Digital, the Diputación General de Aragón-Fondo Social Europeo (DGA-FSE) Grant No. 2020-E21-17R of the Aragon Government, and the European Union, NextGenerationEU Recovery and Resilience Program on 'Astrof\ii sica y F\ii sica de Altas Energ\ii as , CEFCA-CAPA-ITAINNOVA.

\renewcommand{\refname}{References}
\bibliography{refe2}
\bibliographystyle{unsrt}

\end{document}